\documentclass[%
reprint,
superscriptaddress,
 amsmath,amssymb,
 aps,
]{revtex4-1}
\usepackage[pdftex]{graphicx}
\usepackage{dcolumn}
\usepackage{bm}
\usepackage[colorlinks=true,linkcolor=blue]{hyperref}
\usepackage[utf8]{inputenc}
\usepackage[english]{babel}
\usepackage{xcolor}

\newcommand{\un}[1]{\,\mathrm{#1}}

\begin{document}

\preprint{APS/123-QED}

\title{Expansion and fragmentation of liquid metal droplet by a short laser pulse}% Force line breaks with \\

\author{S.~Yu.~Grigoryev}
\email{grigorev@phystech.edu}%Lines break automatically or can be forced with \\
\affiliation{Dukhov Research Institute of Automatics, 22 Sushchevskaya st., 127055 Moscow, Russia}
\affiliation{Landau Institute for Theoretical Physics, RAS, 1-A Akademika Semenova av., 142432 Chernogolovka, Moscow Region, Russia}

\author{B.~V.~Lakatosh}
\affiliation{Moscow Institute of Physics and Technology, 9 Institutskiy per., 141701 Dolgoprudny, Moscow region, Russia}

\author{M.~S.~Krivokorytov}
\affiliation{Institute for Spectroscopy, RAS, Fizicheskaya Street, 5, Troitsk, Moscow, Russia}
\affiliation{EUV Labs, Sirenevy boulevard, 1, Troitsk, Moscow, Russia}

\author{V.~V.~Zhakhovsky}
\email{6asi1z@gmail.com}
\affiliation{Dukhov Research Institute of Automatics, 22 Sushchevskaya st., 127055 Moscow, Russia}
\affiliation{Landau Institute for Theoretical Physics, RAS, 1-A Akademika Semenova av., 142432 Chernogolovka, Moscow Region, Russia}

\author{S.~A.~Dyachkov}
\affiliation{Dukhov Research Institute of Automatics, 22 Sushchevskaya st., 127055 Moscow, Russia}%
\affiliation{Landau Institute for Theoretical Physics, RAS, 1-A Akademika Semenova av., 142432 Chernogolovka, Moscow Region, Russia}
\affiliation{Moscow Institute of Physics and Technology, 9 Institutskiy per., 141701 Dolgoprudny, Moscow region, Russia}
\affiliation{Joint Institute for High Temperatures, RAS, 13/2 Izhorskaya st., 125412 Moscow, Russia}

\author{D.~K.~Ilnitsky}
\affiliation{Dukhov Research Institute of Automatics, 22 Sushchevskaya st., 127055 Moscow, Russia}%

\author{K.~P.~Migdal}%
\affiliation{Dukhov Research Institute of Automatics, 22 Sushchevskaya st., 127055 Moscow, Russia}%

\author{N.~A.~Inogamov}%
\affiliation{Dukhov Research Institute of Automatics, 22 Sushchevskaya st., 127055 Moscow, Russia}%
\affiliation{Landau Institute for Theoretical Physics, RAS, 1-A Akademika Semenova av., 142432 Chernogolovka, Moscow Region, Russia}

\author{A.~Yu.~Vinokhodov}
\affiliation{EUV Labs, Sirenevy boulevard, 1, Troitsk, Moscow, Russia}

\author{V.~O.~Kompanets}
\affiliation{Institute for Spectroscopy, RAS, Fizicheskaya Street, 5, Troitsk, Moscow, Russia}

\author{Yu.~V.~Sidelnikov}
\affiliation{Institute for Spectroscopy, RAS, Fizicheskaya Street, 5, Troitsk, Moscow, Russia}

\author{V.~M.~Krivtsun}
\affiliation{Institute for Spectroscopy, RAS, Fizicheskaya Street, 5, Troitsk, Moscow, Russia}
\affiliation{EUV Labs, Sirenevy boulevard, 1, Troitsk, Moscow, Russia}

\author{K.~N.~Koshelev}
\affiliation{Institute for Spectroscopy, RAS, Fizicheskaya Street, 5, Troitsk, Moscow, Russia}
\affiliation{EUV Labs, Sirenevy boulevard, 1, Troitsk, Moscow, Russia}

\author{V.~V.~Medvedev}
\email{medvedev@phystech.edu}%Lines break automatically or can be forced with \\
\affiliation{Institute for Spectroscopy, RAS, Fizicheskaya Street, 5, Troitsk, Moscow, Russia}
\affiliation{Moscow Institute of Physics and Technology, 9 Institutskiy per., 141701 Dolgoprudny, Moscow region, Russia}%

\date{\today}% It is always \today, today,

\begin{abstract}
We report an experimental and numerical investigation of the fragmentation mechanisms of micrometer-sized metal droplet irradiated  by ultrashort laser pulses. The results of the experiment show that the fast one-side heating  of such a droplet may lead to either symmetric or asymmetric expansion followed by different fragmentation scenarios. To unveil the underlying processes leading to fragmentation we perform simulation of liquid-tin droplet expansion produced by the initial conditions similar to those in experiment using  the smoothed particle hydrodynamics (SPH) method. Simulation demonstrates that a thin heated surface layer generates a ultrashort shock wave propagating  from the frontal side to rear side of the droplet.  Convergence of such shock wave followed by a rarefaction tale to the droplet center results in the cavitation of material inside the central region by the strong tensile stress. Reflection of the shock wave from the rear side of droplet produces another region of highly stretched material where the spallation may occur producing a thin spallation layer moving with a velocity higher than expansion of the central shell after cavitation. It is shown both experimentally and numerically that the threshold laser intensity  necessary for the spallation is higher than the threshold  required to induce cavitation in the central region of droplet. Thus, the regime of asymmetrical expansion is realized if the laser intensity exceeds the spallation threshold. The transverse and longitudinal expansion velocities obtained in SPH simulations of different regimes of expansion are agreed well with our experimental data.
\end{abstract}


\keywords{laser ablation, fragmentation, cavitation, spallation, extreme ultraviolet lithography}

\maketitle

\section*{Introduction}

Fragmentation of liquid droplets underlies a wide range of technological processes including mass spectrometry, liquid fuel dispersion systems, coating deposition, and fabrication of microstructures. In addition, the processes of droplet fragmentation are observed in nature \cite{villermaux:2009} what causes an interest in investigating fragmentation mechanisms. 

An unperturbed liquid droplet is a hydrodynamically stable object. Fragmentation occurs only as a result of a sufficiently strong external influence. Thus, the fragmentation process strongly depends both on the mechanism of an external action and on the properties of a droplet itself. This incorporates a wide range of fragmentation mechanisms which are observed in external electric and magnetic fields \cite{sherwood:1988, eow:2001}, during interaction with a gas jet \cite{guildenbecher:2009}, in collisions with a solid obstacle \cite{villermaux:2011}, in collisions of droplets with one another \cite{pan:2005}, or under the action of a laser pulse \cite{klein:2015}. 

The most intense fragmentation occurs under the influence of shock waves. A shock wave can propagate in an extrenal environment~\cite{boiko:2007,boiko:2012} or be formed within a droplet by applying an external force on a time scale substantially shorter than the time of sound propagation through it. The latter occurs when droplets collide with a solid surface at very high velocities~\cite{field:1989} or when a droplet is irradiated by short laser pulses~\cite{stan:2016,lindinger:2004,krivokorytov:2017,basko:2017,vinokhodov:2016,krivokorytov:2018}. Laser energy can be released both in the volume of a droplet~\cite{stan:2016,lindinger:2004} and at its surface~\cite{krivokorytov:2017} what depends on the optical properties of a material at a laser wavelength. In the last case the fragmentation scenario may vary.

The problem of interaction of a laser pulse with a droplet has attained practical interest during the development of extreme ultraviolet (EUV) sources used in the next-generation industrial lithography~\cite{fujimoto:2012,fomenkov:2017}. It was found that irradiation of a tens-micrometer sized liquid-metal tin droplet by two successive laser pulses produces plasma which emit photons in the EUV range~\cite{fujimoto:2012,fomenkov:2017,fujioka:2008}. The first pulse (known as the ``pre-pulse'') serves to optimize the target by deforming~\cite{kurilovich:2016} or even fragmenting it~\cite{fujioka:2008}. The second pulse (known as the ``main pulse'') heats the material to a high-temperature plasma state. Fujimoto et al.~\cite{fujimoto:2012} demonstrated that the highest efficiency of the main pulse energy conversion into the EUV is achieved with the use of picosecond pre-pulses. 

Our previous work~\cite{krivokorytov:2017,vinokhodov:2016} presented a phenomenological description of fragmentation of liquid-metal droplets exposed to Ti:sapphire laser pulses. We also discussed possible physical fragmentation mechanisms associated with the propagation of shock waves in spherical samples~\cite{basko:2017}. In the present paper, we report the systematic study of a droplet response to short laser pulses of various energies (intensities). Using the method of instantaneous shadow photography, we observe that a droplet subjected to laser pulses undergoes a strong expansion with the formation of internal cavities. This means that a uniform droplet takes the form of a soap bubble with a liquid-metal shell. At high intensities the formation of two cavities is clearly observed: one (front) cavity is closer to the side of a droplet irradiated by a laser pulse, and the other (rear) cavity appears at the opposite side. When the laser intensity decreases, the expansion rate of the rear cavity falls considerably faster than that of the front cavity. At the certain intensity the rear cavity disappears completely. In addition, a variation in the intensity of laser pulses affecting a droplet changes the fragmentation scenario qualitatively. At high intensities a liquid-metal shell is fragmented during expansion. At low intensities, when the rear cavity disappears, the expansion of a shell becomes limited, and it begins to shrink due to surface tension. In this case, the shell finally breaks apart on internal nonuniformities what results in the formation of jets and smaller droplets. 

To understand the mechanism of the cavities formation within a droplet, we perform the detailed numerical simulations of a shock waves propagation induced by a laser pulse. The smoothed particle hydrodynamics (SPH) method is used as the most appropriate one to model flows with the loss of continuity and fragmentation. The results of the numerical study clearly demonstrate that the formation of cavities is a result of exceeding liquid tensile strength under the action of strong rarefaction stresses which follow the shock front. 

%\section*{Experimental results}

\section*{Experimental setup}

\begin{figure}[t]
	\flushleft{
		\includegraphics[width=1.0\linewidth]{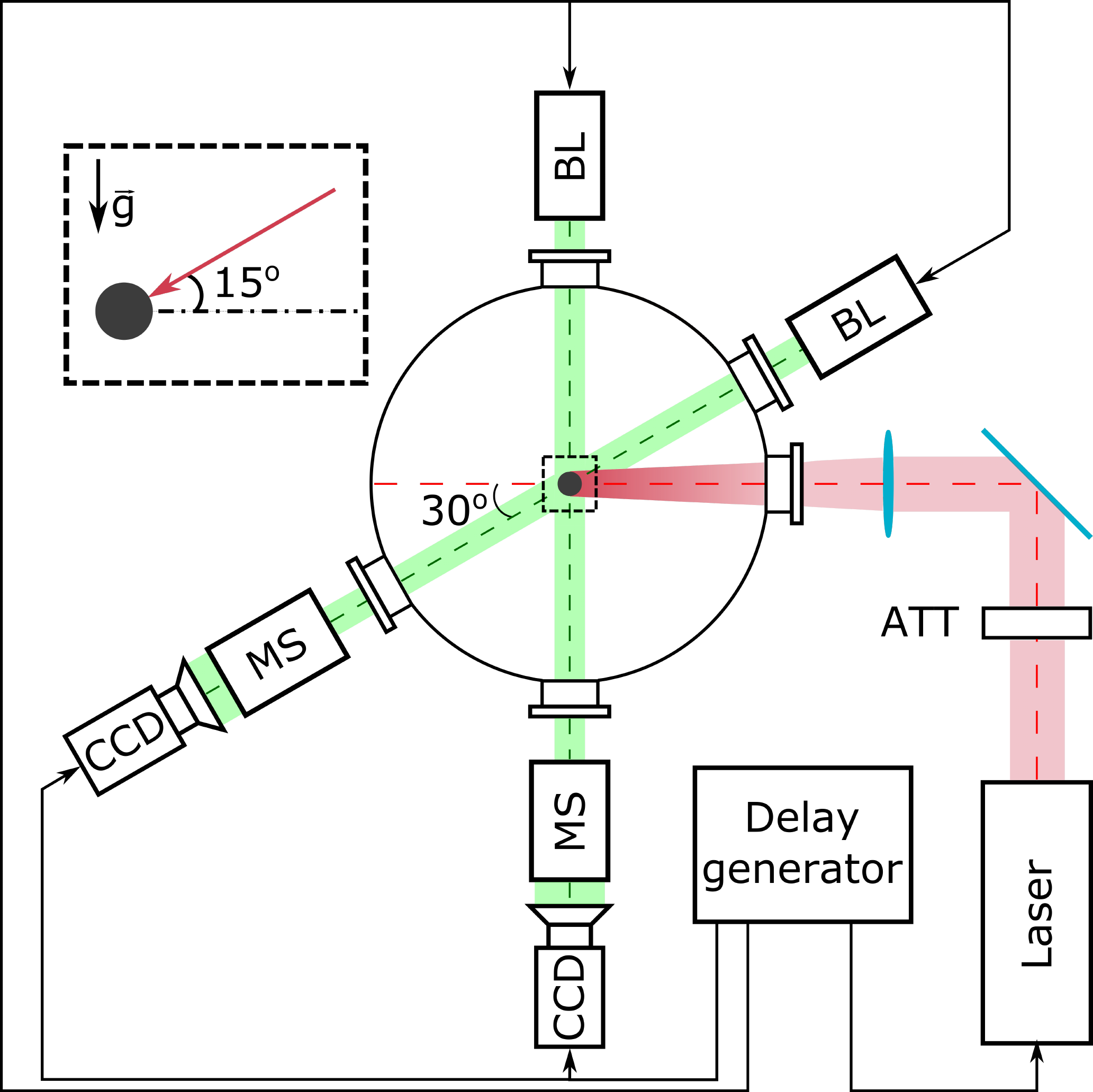} \\
		\caption {\label{fig:facility} The experimental setup scheme to study the liquid-metal droplet fragmentation by short laser pulses: (Laser) Ti:sapphire laser; (CCD) CCD camera; (MS) microscope; (ATT) attenuator; (BL) pulsed back illumination. Droplets move perpendicular to the plane of the figure. The Ti:sapphire laser beam comes at 15$^\circ$ to the plane of the figure.}
	}
\end{figure}

The experimental setup scheme is shown in Fig.~\ref{fig:facility}. The experiment is performed in the vacuum chamber where the residual gas pressure is less than~$10^{-2}\un{Pa}$. The Sn--In liquid-alloy targets (in a mass proportion of 52\%--48\%) are used instead of the pure tin ones to decrease the melting temperature from $232^\circ\un{C}$ (Sn) to $119^\circ\un{C}$ (Sn--In). That greatly simplifies the experiment while tin component provides the required plasma for the EUV light generation. To produce targets we use the previously developed droplet generator~\cite{vinokhodov:2016} ensuring the Plateau--Rayleight instability to split jets into droplets. The generator is synchronized with laser pulses what provides an accurate irradiation of a target. The laser beam is focused on a free-flying drop of liquid metal at a sufficient distance from the generator nozzle to guarantee the natural oscillations of a droplet are damped. The diameter of the droplets generated throughout the experiments is constant and amounts to $49.0 \pm 1.6\un{\mu m}$.

Figure~\ref{fig:facility} shows that the incident laser beam is~$15^\circ$ with the horizontal plane. We use the Ti:sapphire laser ($\lambda = 780-820\un{nm}$) with the fixed pulse duration $\tau_L = 800\un{fs}$ and the Gaussian intensity profile in the focal plane. The size of the focal spot defined as the full width at the half-height is $D_L = 60\un{\mu m}$. The laser pulse energy ($E_L$) varies in the experiments from 0.08 to 1.66 mJ. The corresponding intensity of the laser pulses, which is defined as $I_L = 4E_L/(\tau _L \pi D_L^2)$, varies in the range~$(0.4-8.0) \times 10^{13}\un{W/cm^2}$.

The droplet response to the focused laser pulse is recorded using the shadow photograph method. CCD cameras are directed perpendicular to a drop fall with the $30^\circ$ angle to the horizontal projection of the laser beam (Fig.~\ref{fig:facility}). To resolve small droplets the cameras are equipped with microscopes. The 30-ns pulsed laser is mounted opposite each camera for back illumination what determines exposure. The droplet velocity is about 10 m/s, so that images can be considered instantaneous with a good accuracy. By varying the delay between the camera and the Ti:sapphire laser pulse, it is possible to obtain images of the droplet shape evolution at its various stages. To ensure accuracy we perform measurements with 100 samples.

\begin{figure}[t]
	\center{
		\includegraphics[width=1.0\linewidth]{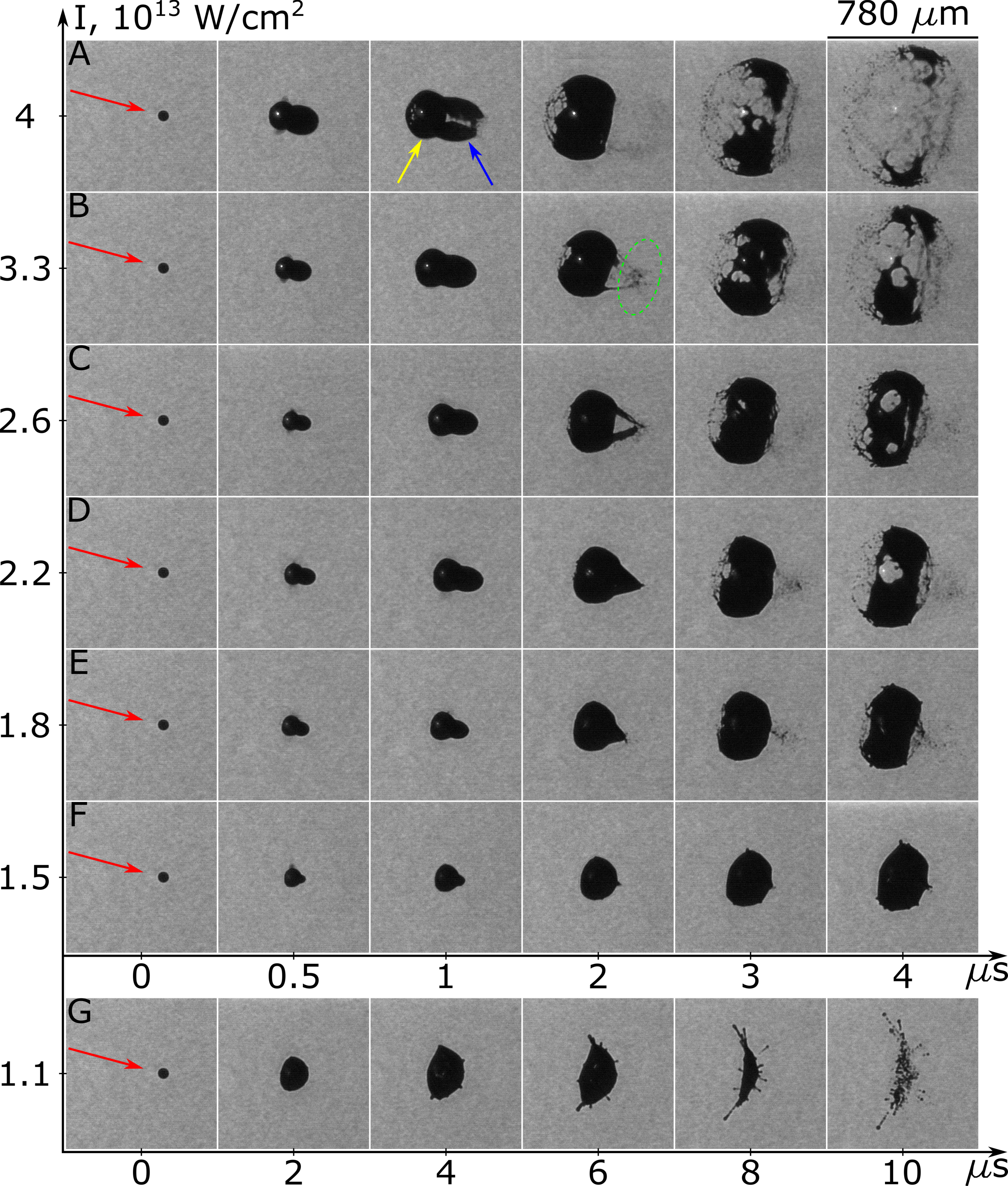} \\
		\caption{Side views of target shape evolution for different laser pulse intensities presented on the vertical axis. The horizontal axes shows the time delays relative to the laser pulse. The red arrow indicates the direction of laser beam at an angle of $30^\circ$; the yellow and blue arrows show the central and rear-side shells, respectively. A green ellipse identifies a cloud of smaller droplets formed after the fragmentation of the rear-side shell.}\label{fig:drop_decreas_int}
	}
\end{figure}

\section*{Evolution of droplet shape observed in experiments}

The obtained shadowgraphs of the time evolution of liquid-metal droplets, which are subjected to short laser pulses, are ordered by the applied laser intensity $I_L$ (Fig.~\ref{fig:drop_decreas_int}). The range of laser intensities varies from $1.1\times10^{13}$ to $4.0\times10^{13}\un{W/cm^2}$. The observed expansion and fragmentation process of a liquid-metal droplet qualitatively depends on a laser pulse intensity focused on it. 

Our detailed analysis begins with the droplet response to the most intense laser pulse with $I_L=4\times10^{13}\un{W/cm^2}$ shown in Fig.~\ref{fig:drop_decreas_int}A. One should notice that the evolution of the droplet shape subjected to a more intense laser pulses $I_L \ge 4 \times 10^{13}\un{W/cm^2}$ has not new qualitative features except for an increase in the expansion rate. This scenario persists even with a slight change in the other parameters of the experiment, i.e. the size of the droplet, the size of the laser focal spot, etc. A similar regime of the liquid-metal droplet response to the action of a femtosecond laser pulse has already been described in detail in our previous paper \cite{krivokorytov:2017}. Fig.~\ref{fig:drop_decreas_int}A shows that after the action of the laser pulse the droplet asymmetrically expands over time, so that the droplet volume increases tenfold in 500\,ns. As the mass of the droplet conserves, this expansion can be explained by the formation of two cavities within the droplet. Below, the first cavity arising in the center of the droplet is surrounded by the front shell (yellow arrow in Fig.~\ref{fig:drop_decreas_int}A) and the second cavity emerging at the back side of the surface is surrounded by the rear shell (blue arrow in Fig.~\ref{fig:drop_decreas_int}A). It is seen that both shells expand and eventually fragment over time. However, the rear shell fragments faster than the front shell. Thus, $2\un{\mu s}$ after the action of the laser pulse, the rear shell is completely fragmented, whereas the front shell still remains intact.

\begin{figure}[t]
	\centering{
	\includegraphics[width=1.0\linewidth]{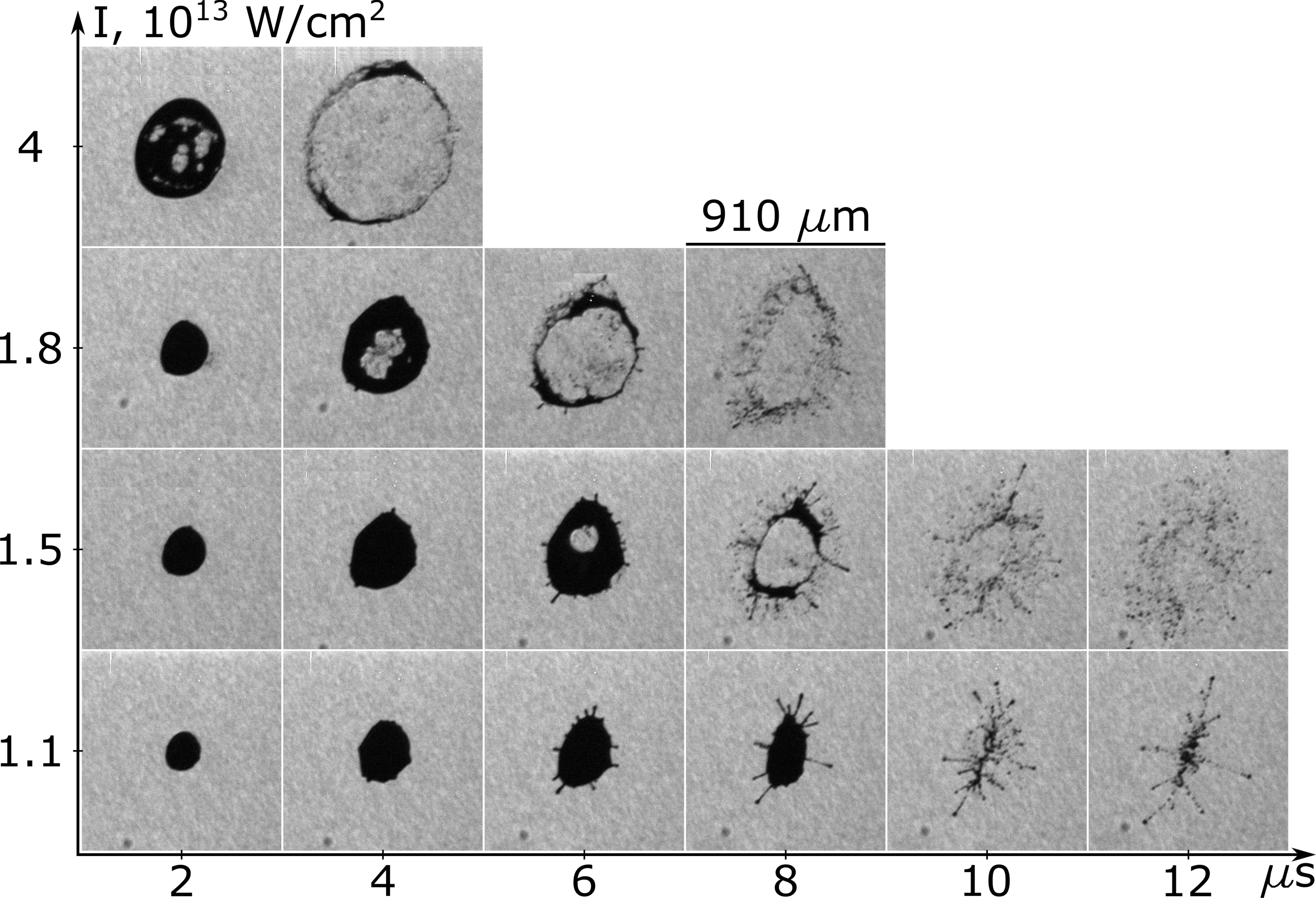}
	\caption{Rear-side views of the target shape evolution for different laser pulse intensities presented on the vertical axis. The horizontal axes indicate a frame delay relative to the laser pulse.}
	\label{fig:Drop_low_int}
	}
\end{figure}

A decrease in the laser pulse intensity leads to a decrease in the shells expansion rate and a change in their geometry. We noticed that under irradiation with $I_L\ge4\times10^{13}\un{W/cm^2}$ the rear-side shell expands until fragmentation. However, with the decrease in intensity the shape of the rear shell gradually changes from the elongated ellipsoid into the cone-like form (Figs.~\ref{fig:drop_decreas_int}B--\ref{fig:drop_decreas_int}E). Finally, the rear shell also becomes fragmented, but the process develops in a different scenario. The collapse occurs due to the shell narrowing around the laser pulse direction, not to the expansion.

The decrease in the laser intensity down to $I_L=1.5\times10^{13}\un{W/cm^2}$ (Fig.~\ref{fig:drop_decreas_int}F) leads to the considerable shrink in the rear shell size, the further intensity decrease to $I_L=1.1\times10^{13}\un{W/cm^2}$---to the rear shell disappearance~(Fig.~\ref{fig:drop_decreas_int}G). Nevertheless, the front shell is formed, but under the laser energy decrease its expansion becomes limited: after $\sim6\un{\mu s}$ from the laser irradiation the shell also begins to collapse.

It is also worth noting that at low intensities, starting at~$\sim4\un{\mu s}$, convexities appear on a droplet surface which then evolve into jets. These convexities are observed even at $I_L \le 1.5 \times 10^{13}\un{W/cm^2}$~(Figs.~\ref{fig:drop_decreas_int}F, \ref{fig:drop_decreas_int}G). The morphology of the expansion of fragmentation products in later times also changes: it becomes flatter in comparison to the results obtained at high-intensity laser pulses (Figs.~\ref{fig:drop_decreas_int}A, \ref{fig:drop_decreas_int}G). 

A change in the morphology of the fragmented droplet at low laser intensities, which is the damping of the rear shell formation, is also observed in the projection perpendicular to the laser beam. Figure~\ref{fig:Drop_low_int} shows that for any laser intensity the expansion of a droplet is symmetric in this projection at earlier stages. But the later fragmentation scenario is changing. At $I_L \ge 1.8 \times 10^{13}\un{W/cm^2}$ we observe the formation of a ring; at $I_L = 1.5 \times 10^{13}\un{W/cm^2}$ jets are formed on the ring; and at $I_L = 1.1 \times 10^{13}\un{W/cm^2}$ the ring does not appear at all: only jets are observed. 

\begin{figure}[t]
\center{
\includegraphics[width=1.0\linewidth]{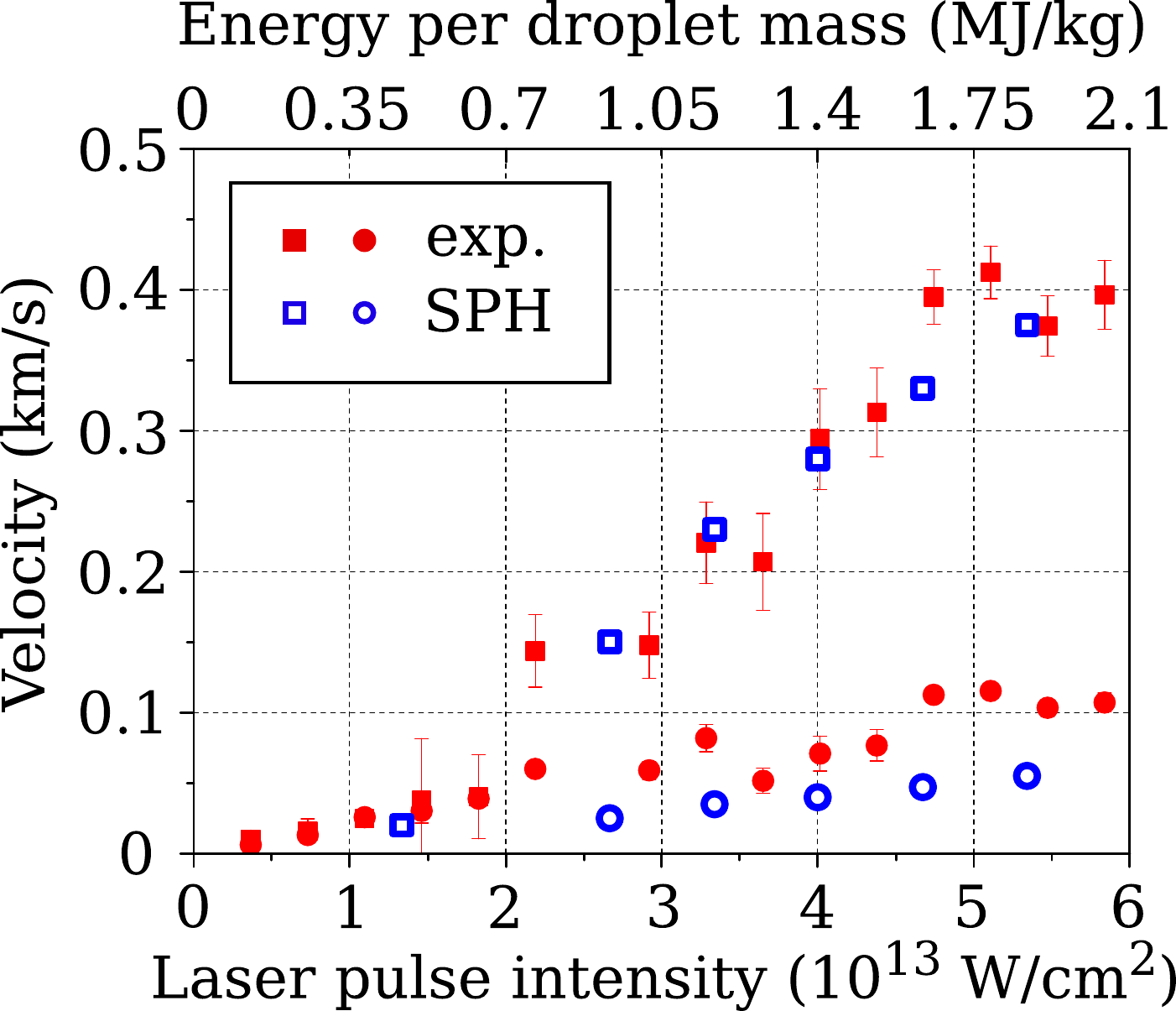}
\caption {\label{fig:vel} Expansion velocities of the rear (squares) and spherical (circles) shells as functions of pulse intensity (low axis) or exposure energy per droplet mass (top axis). Red symbols correspond to the experimental data, and the blue ones show simulation data. The top axis is linked to the low one via the absorption coefficient of 0.125.}
}
\end{figure}

At late stages the dispersed fragments of a droplet continue to move by inertia.
Their size distribution lies within a range of several micrometers or less. The velocity of fragments movement can be estimated from the expansion rates of the shells in the corresponding directions. The results of measuring the characteristic expansion rates of the shells along and perpendicular to the laser beam are shown in Fig.~\ref{fig:vel}. One can notice that the shell expansion rates depend on the laser pulse intensity. It is interesting that the expansion rate of the rear shell along the laser beams reaches 500\,m/s at $I_L = 7.3 \times 10^{13}\un{W/cm^2}$, while the expansion rate of the front shell perpendicular to the laser beam is much less at the same intensity and reaches only 130\,m/s. Fig.~\ref{fig:vel} also shows that at $I_L \le 1.1 \times 10^{13}\un{W/cm^2}$ the front and rear shells expansion rates coincide what indicates the rear shell disappearance. 

\section*{Simulation setup}

\subsection*{Equations of material motion}

The simulated tin droplet is a sample represented by a continuous compressible material. Its evolution is determined by the mass, momentum, and energy conservation equations: 
\begin{gather}
\label{eq:continuity}
\dot{\rho} + \rho\, \nabla \cdot \mathbf{U}=0, \\
\label{eq:momentum}
\rho \dot{\textbf{U}} + \nabla P=0, \\
\label{eq:energy}
\rho \dot{E} + \nabla \cdot \left(P \mathbf{U}\right) = Q,
\end{gather}
where $\rho$ and $\mathbf{U}$ are the density and the velocity of the material respectively; $P$ is the hydrostatic pressure, which is determined from the equation of state; $E=e+\mathbf{U}^2/2$ is the total specifc energy which consists of the internal and kinetic specific energies; and $Q$ is the volumetric heat source which emulates the heating induced by laser radiation.

The reduced system of Eqs.~(\ref{eq:continuity})--(\ref{eq:energy}) completed by the equation of state is solved during simulation of tin droplet expansion and fragmentation by laser pulses using the SPH method. Its basic idea is to represent a continuous medium by a set of lagrangian particles. In contrast to the conventional grid methods the governing equations are represented in integral form, after which a transition is performed from integration to summation over neighboring SPH particles within the ``smoothing radius''. The usage of Riemann solver at an inter-particle contact \cite{monaghan:1997,parshikov:2002} to obtain a contact  pressure and velocity gives a good accuracy in simulation of shock wave phenomena. Moreover, simulation of discontinuities, which occur under the influence of tensile stresses in material, is performed in a natural way without special algorithms for handling complex boundaries. The movement of free external and internal boundaries and surfaces can be directly tracked and compared to the experimental ones.

\subsection*{Equation of state for liquid tin }

The system~(\ref{eq:continuity})--(\ref{eq:energy}) is enclosed with an equation of state which couples the hydrostatic pressure, the density, and the interntal energy. Unlike the experiment we use a pure tin droplet in simulation instead of a liquid-metal tin--indium alloy. Indium and tin have close thermomechanical parameters: the difference in the liquid indium and tin densities is less than 1\% ($7030\un{kg/m^3}$ and $6980\un{kg/m^3}$ respectively) \cite{assael2010reference, assael2012reference}; the sound velocities are $2.59$ (In) and $2.49$ (Sn) km/s \cite{1980:LASL.Shock}; the surface tension of these materials, which determines the residual properties, is almost the same and equals to $0.55\un{N/m}$ \cite{kononenko1972surface}. Thus, all the features of a simulated shock propagation will remain.

We use the Mie--Gr$\mathrm{\ddot{u}}$neisen equation of state for liquid tin:
\begin{equation}
\label{eq:mie-grun}
P = P(\rho, e) = P_{r} + \Gamma \rho (e - e_{r}),
\end{equation}
where $\Gamma$ is the Gr$\mathrm{\ddot{u}}$neisen parameter, and $P_r$ and $e_r$ are the reference pressure and specific energy represented by the shock hugoniot. The latter is usually defined using the linear approximation for the shock velocity $u_s$ dependence on the material velocity $u_p$: $u_s = c + su_p$, where $c$ is the bulk sound velocity in normal state and $s$ is the parameter. Thus, the reference pressure and specific energy curves take the form: 
\begin{equation}
\label{eq:eos_ref}
P_{r}(\rho) = \rho_0c^2 \frac{1-x}{[1-s(1-x)]^2}, \qquad e_{r} = \frac{P_{r}}{\rho} \frac{1-x}{2},
\end{equation}
where $x = \rho_0 / \rho$ is the reversed compression ratio. The values for the liquid tin equation of state parameters are given in Table~\ref{table:tin}.

\begin{table}[t]
\center{
\caption{\label{table:tin} Properties of liquid tin. }
\label{tab:a}
\tabcolsep10pt
\begin{tabular}{ll}
\hline
\hline
    {Mechanical properties}                                  & {Value}     \\
\hline
Initial density $\rho_{0},\,\mathrm{kg/m}^{3}$                & $6824$       \\
Compression modulus $B,\,\mathrm{GPa}$                        & $37.7$       \\
Specific heat $C_v,\,\mathrm{J/(kgK)}$                         & $227$        \\
\hline
\hline
    {Equation of state parameters}                            &              \\
\hline
Parameter $\Gamma$                                            & $1.486$      \\
Parameter $c,\,\mathrm{km/s}$                                 & $2.45$       \\
Parameter $s$                                                 & $1.45$       \\
\hline
\hline
\end{tabular}
\tabcolsep15pt
}
\end{table}

\subsection*{Tensile strength of liquid tin}

Our initial simulations demonstrated that the propagation of the shock induced by laser irradiation forms a condition for the cavity formation within the droplet due to high tensile stresses which follow the shock. The cavitation condition, as well as the spallation, is the exceeding of tensile strength within material. At this point the medium reacts by forming free internal boundaries. The choice of this terminology does not contradict the generally accepted definitions of these concepts and is used to explicitly separate the studied processes. Otherwise, we mean that the processes of cavitation and spallation have a single physical nature associated with stress relaxation in response to the stretching of the material.

The critical value of tensile stresses (or tensile strength) is not a property of the material in question, but depends on the nature of the loading, in particular on the strain rate. The information about the tensile strength in liquid tin is rather scarce. It is known that for most metals the tensile strength in a material is greatly reduced at a transition from solid to liquid. Thus, the tensile strength in tin decreases by an order of magnitude: from $1.2\un{GPa}$ in the solid state to $0.12\un{GPa}$ in the molten state \cite{kanel:2015}. However, the tensile strength increases with the strain rate: the experiment of Ashitkov et al. shows that the tensile strength in tin is 1.9 GPa at the strain rate of $1.3\times10^{9}\un{s}^{-1}$ \cite{ashitkov:2016}.

\begin{figure}[t]
\center{
\includegraphics[width=1.0\linewidth]{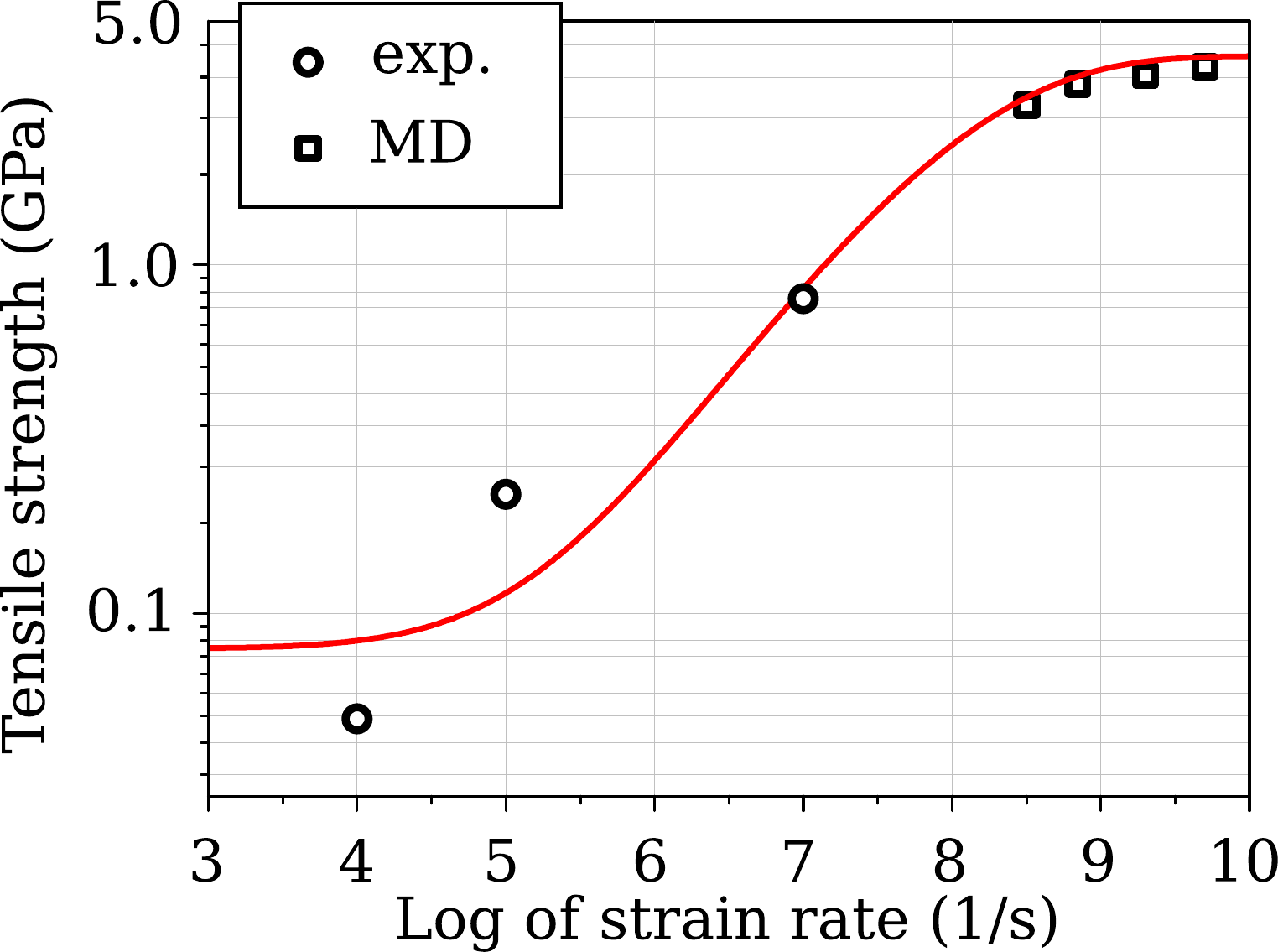} \\
\caption {\label{fig:tensile_strength} Tensile strength in liquid tin vs. the strain rate. The solid line is fitted to experimental data \cite{grady1988spall, kanel1996spall, de2007experimental} (circles) and the molecular dynamic simulations (squares).}
}
\end{figure}

To take into account the dependence of the tensile strength on the strain rate, fittings are made from known experimental points and molecular dynamics simulations (Fig.~\ref{fig:tensile_strength}). The obtained tensile strength dependence on the strain rate is used in SPH simulations as the criterion under which cavitation occurs. The implementation of the criterion is that each pair of particles, for which the Riemann problem solution reaches or exceeds the tensile strength at the current strain rate, loses connectivity with each other. This condition is accompanied by an instant relaxation of tensile stresses. 

\subsection*{Heating of surface layer}

\begin{figure*}[t]
\center{
\includegraphics[width=0.9\linewidth]{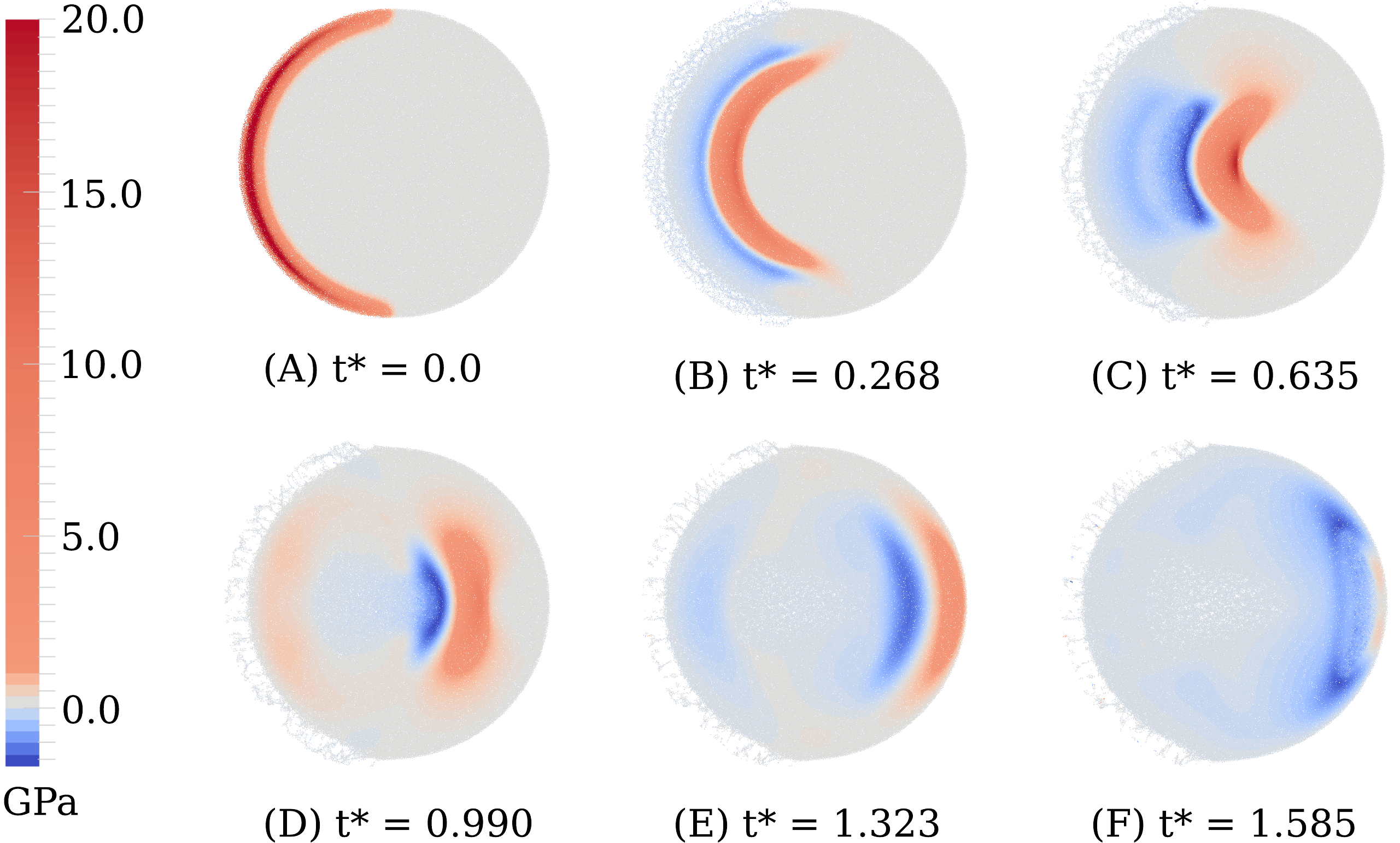} \\
\caption {\label{fig:2E0pres} Pressure maps at successive instants of time ($t^{*} =  ct/R$): (A) the initial shape of the pressure pulse (maximum amplitude at the frontal focus, minimum amplitude near the equator); (B) the decrease in the amplitude of the pressure pulse and its broadening during its propagation into the droplet; (C) the growth of the amplitude of the compression and stretching waves due to the focusing effect, the formation of cavitation bubbles in the center of the droplet; (D) the decrease in the amplitude of the compression wave after passing through the center of the droplet, further formation of cavitation bubbles; (E) the amplitude of the tensile stresses becomes less in absolute value than the spall strength of tin, the formation of cavitation bubbles at the center of the droplet stops; (F) the formation of a spall at the posterior pole of the droplet. The laser pulse propagates from left to right.}
}
\end{figure*}

The energy of the femtosecond laser pulse is absorbed by free electrons in metals during a period that does not exceed $\sim100\un{fs}$ ($\sim10\un{fs}$ for tin) what may be considered instantaneous in contrast to the 800-fs laser pulse. Electron-ion energy transition occurs due to the collisions between heated electrons and cold ions. The electron-ion relaxation finishes within $\sim10\un{ps}$ what results in an equilibrium temperature within the skin layer of a droplet. Our simulations do not take into account these relaxation processes: the laser energy absorption is modeled by setting the resulting intertnal energy within the skin layer. The source term in the energy transport equation \eqref{eq:energy} is given by:
\begin{equation}
\label{eq:energI-3D}
Q = E_{in,0}\exp\left(-\left(\frac{r-R}{\delta R} \right)^2 \right) H\left(\pi / 2 - \theta\right)\cos\theta.
\end{equation}
Here $E_{in,0}$ is the aborbed energy amplitude, $R$ is the radius of the droplet, $r$ is the distance from the center of the droplet, $\delta R$ is the skin layer depth at which the absorbed energy is reduced by $e$ times ($70\un{nm}$), and $\theta$ is the angle between the laser beam and a point within the droplet. 

The skin layer depth $\delta R$ of initial heating is estimated using the two-temperature (2T) hydrodynamic model of a nonequilibrium system of electrons and ions. The calculation is carried out using electron--ion exchange coefficients and electronic thermal conductivity obtained from ab-initio calculations of liquid tin at various electronic temperatures. The obtained equilibrium temperature profile takes a gaussian shape with the corresponding skin layer depth about $60$--$80\un{nm}$ in a wide range of absorbed fluences.

The angular dependence of the absorbed energy $Q(\theta)$ in \eqref{eq:energI-3D} describes the inhomogeneous character of the droplet heating. The absorbed energy varies depending on the incidence angle due to the conditions for a laser beam reflection. It takes a maximal value when the plane surface is irradiated at $\theta = 0$ and decrease at $\theta > 0$ if one ignores light polarization. A proper incident angle dependency for spherical geometry is the cosine function. Finally, the Heaviside function $H(\pi / 2 - \theta)$ determines the one-sided laser radiation.

\begin{figure*}[t]
\center{
\includegraphics[width=0.9\linewidth]{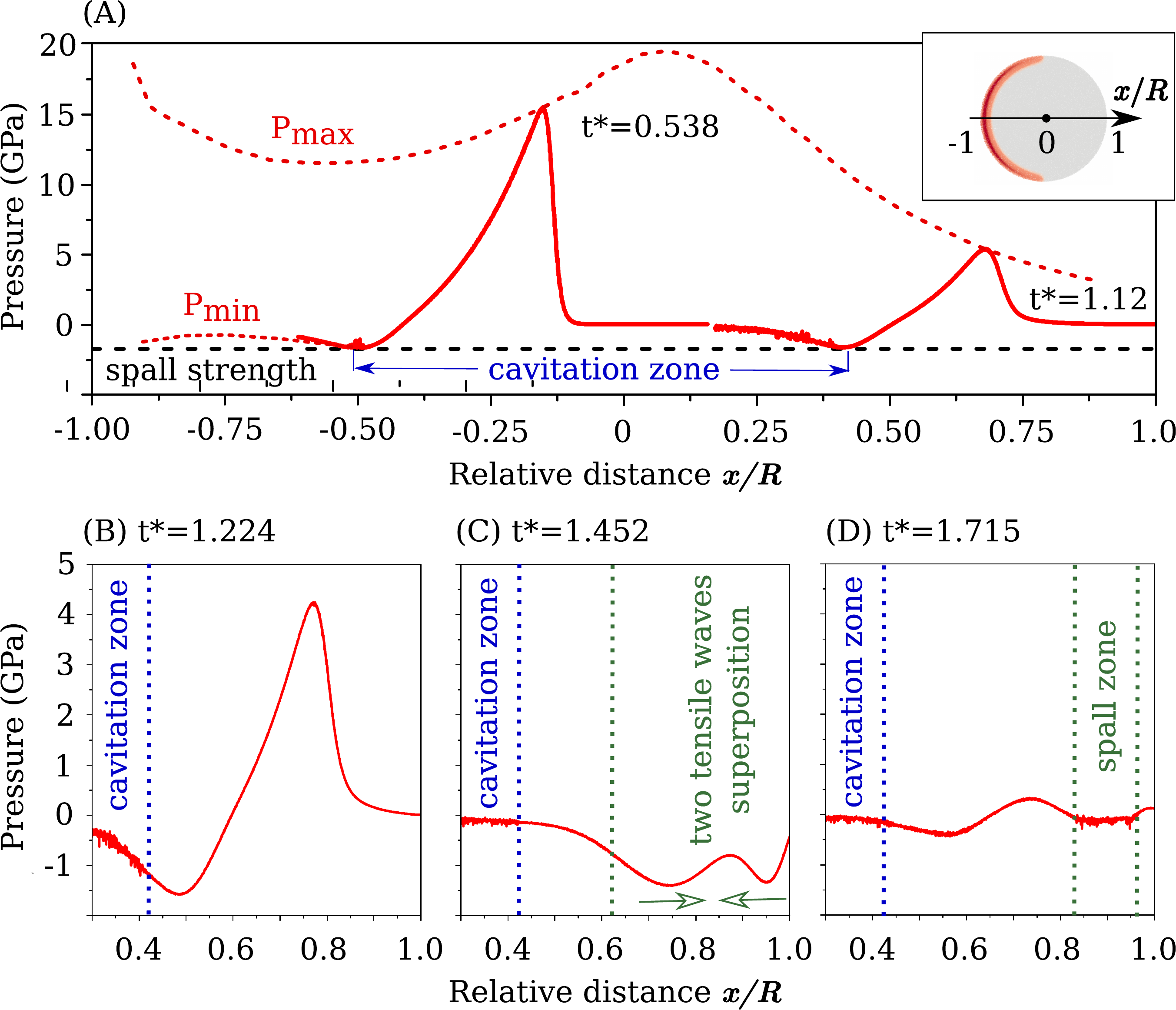} \\
\caption {\label{fig:profile} Evolution of the pressure profile along the axial line: (top) the formation of the cavitation bubble in the cavitation zone $x/R=(-0.5,0.4)$ as a result of the focusing effect; (bottom) the formation of a spall at the posterior pole of the droplet as a result of superposition of two tensile waves.}
}
\end{figure*}

The direct 3D simulation of droplets with experimental sizes is resource consuming since the skin layer and the droplet radius differ by almost 3 orders of magnitude ($R\sim25\un{m}$, $\delta R\sim100\un{nm}$). We need at least 5-10 SPH particles to resolve the skin layer what results in $~\sim10^{11}$ SPH particles for the droplet. Therefore, in the present paper we confine ourselves to smaller droplets ($R=1,\,2\un{\mu m}$). The possible scaling effects are discussed in the next Section.

\section*{Simulation results}

%\subsection*{General phenomenology}

The detailed analysis of processes within the irradiated liquid-metal drop with the diameter $D = 2\un{\mu m}$ is based on SPH simulation results. The dynamic response of the droplet to the high-intensity experimental regime is illustrated in Fig.~\ref{fig:2E0pres} by series of 2D pressure maps. The absorption of the laser energy leads to the almost instantaneous heating of the skin layer with thickness of $\delta R\sim100\un{nm}$ which is accompanied by an increase in pressure up to $\sim30\un{GPa}$. One should notice that due to the introduced heating inhomogeneity $Q(\theta)$ in \eqref{eq:energI-3D} the pressure near the equatorial zone is reduced to zero (Fig.~\ref{fig:2E0pres}A). Release at the front surface leads to the skin layer ablation induced by the material transport in a direction perpendicular to the droplet surface. Subsequently, the ablated material is of no interest to us; therefore, we neglect the ablated particles in calculations for the convenience of presenting the main results.

\begin{figure*}[t]
\center{
\includegraphics[width=0.9\linewidth]{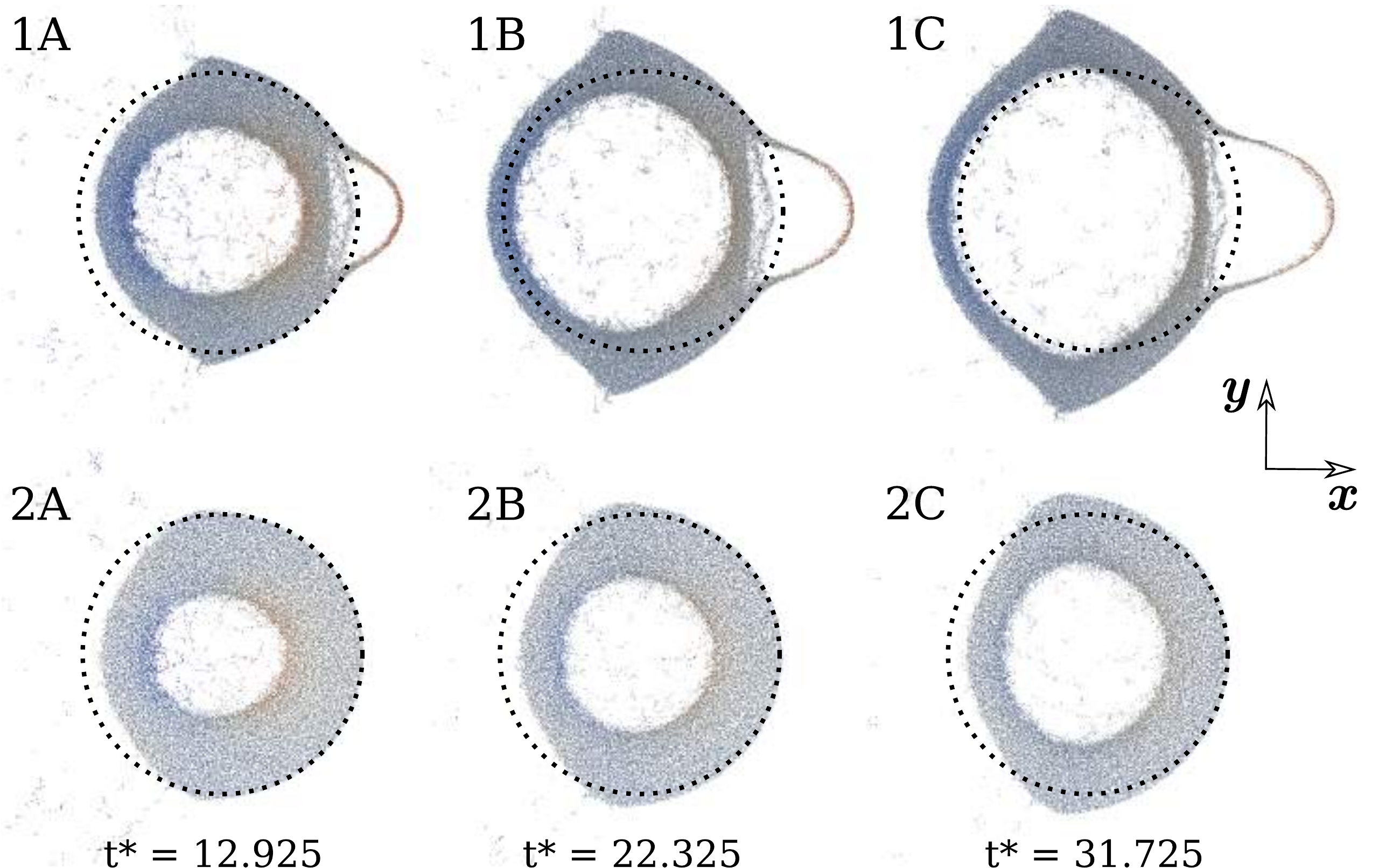} \\
\caption {\label{fig:frag} Two regimes of droplet evolution: (A) high-intensity regime (the growth of a massive cavitation bubble in the center of the droplet is observed, as well as a spall at the posterior pole); (B) moderate-intensity regime (a rear-side spallation at the posterior pole of the droplet is absent, while the growth rate of the cavitation bubble in the center decreases).}
}
\end{figure*}

The pressure pulse, which is formed due to irradiation, begins to converge to the center of the droplet. The tensile wave, which radial profile has a typical triangular form, moves behind it. The initial pressure pulse is very short and comparable with the skin layer depth, so its propagation is accompanied by a quite rapid decrease in its amplitude and spatial broadening \cite{grady1996high}. Thus, at a distance of $x/R\sim0.2$ from the surface, the amplitude of the compression wave decreases by a factor of $\sim2$ and continues to fall gradually to $12\un{GPa}$ at a distance of $x/R\sim0.5$ (Fig.~\ref{fig:profile}A). The tensile wave is characterized by a similar trend. Further in this paper we use dimensionless quantities of length and time: the distance from the droplet center along the axial line is reduced to the droplet radius $x^{*}=x/R$, and the dimensionless time $t^{*}$ is expressed through the ratio of the droplet radius to the speed of sound $t^{*}= ct / R$. Visually, the pulse broadening and the decrease in its amplitude are observed in the reduced pressure maps (Fig.~\ref{fig:2E0pres}B).

The pulse convergence towards the center of the droplet is accompanied by the gradual decrease in its amplitude what is the result of the focusing effect. However, at a certain moment the amplitude begins to increase as shown in Fig.~\ref{fig:profile}A. The phenomenon occurs due to the localization of the pulse energy within a smaller volume near the droplet's center. As a result, the specific energy per unit volume increases what leads to the rise in pressure. Quantitatively, the shock wave amplitude increases from $12\un{GPa}$ at a distance $x/R=0.5$ from the irradiated surface to $20\un{GPa}$ near the center of the droplet (Figs.~\ref{fig:2E0pres}C and \ref{fig:profile}A). The focusing effect is valid not only for compression, but also for tensile stresses (Figs.~\ref{fig:2E0pres}C and \ref{fig:profile}A). The amplitude of the tensile wave increases in absolute value and at a distance $x/R=0.5$ from the irradiated surface it reaches the tensile strength in tin. Relaxation of critical tensile stresses results in the formation of multiple cavitation bubbles which coagulate into a single cavity in the center of the droplet (Fig.~\ref{fig:frag}). Figure~\ref{fig:profile}A shows the onset of the cavitation bubbles formation (cavitaion zone). 

\begin{figure*}[t]
\center{
\includegraphics[width=1.0\linewidth]{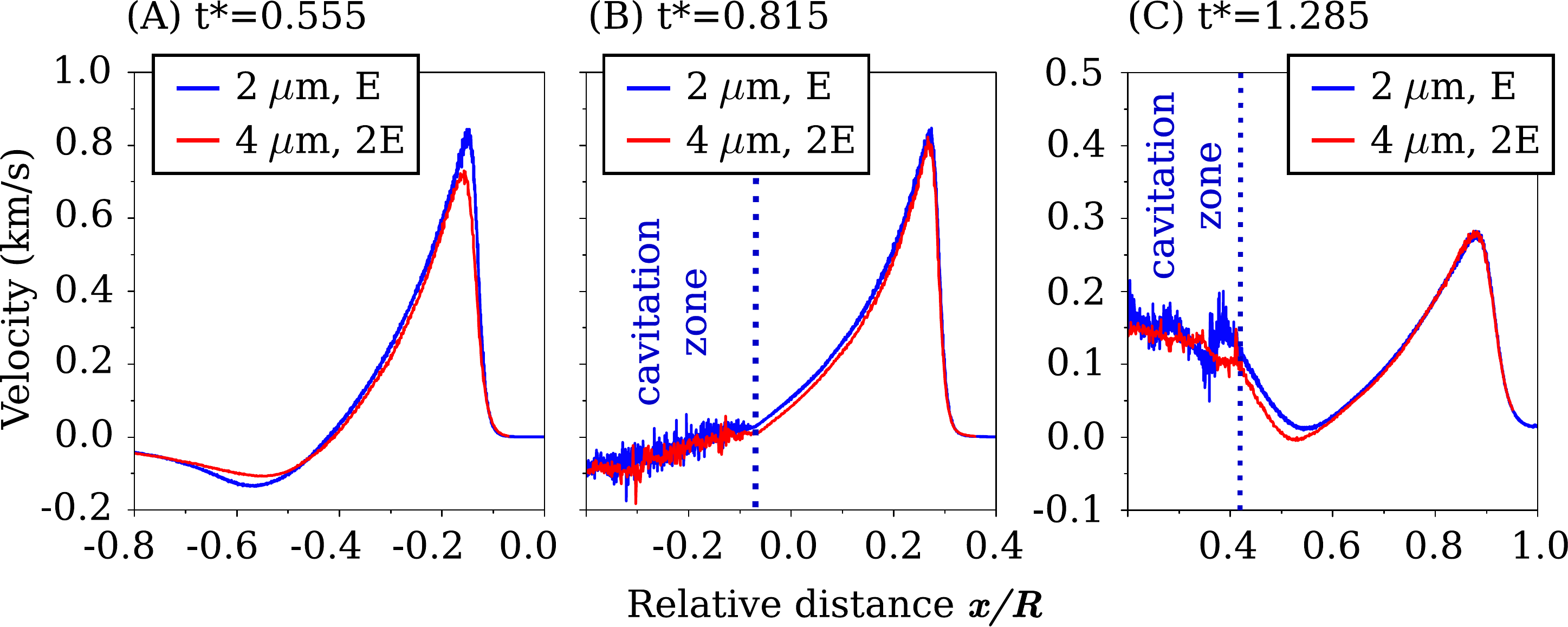} \\
\caption {\label{fig:vx-spall} Velocity profiles along the axial line (left) before the formation of cavitation bubbles, (center) at the time when the pulse passes through the center of the droplet, and (right) at the moment of arrival of the pulse to the rear side. The comparison is given for droplets with diameter $D = 2\un{\mu m}$ (blue curve) and $D = 4\un{\mu m}$ (red curve). The distance is normalized to the radius of the droplets.}
}
\end{figure*}

Focusing of the pulse energy within the droplet's center is followed by the shock wave divergence with a decrease in its amplitude (Figs.~\ref{fig:2E0pres}D, \ref{fig:profile}A). In particular, the amplitude decreases by an order of magnitude from $20\un{GPa}$ in the center of the droplet to $\sim3\un{GPa}$ at the moment when the shock wave approaches the opposite side (Figs.~\ref{fig:2E0pres}E,~\ref{fig:profile}A,~and~\ref{fig:profile}B). The decrease in an amplitude due to a wave divergence is also valid for tensile waves. The tensile stresses become lower in an absolute value at the distance $x/R=0.4-0.5$ than the tensile strength in tin what ceases the formation of cavitation bubbles (Figs.~\ref{fig:2E0pres}E, \ref{fig:profile}A). 

The shock wave profile takes the triangle form for positive pressures at the moment of the arrival to the opposite side of the droplet (Fig.~\ref{fig:profile}B). The amplitude is $\sim 3\un{GPa}$ and decreases linearly behind the front. The reflection of the pressure pulse from the free boundary induces the release with the backpropagation of the additional tensile wave. The spallation condition is satisfied near the opposite pole as a result of superposition of two tensile waves (Figs.~\ref{fig:profile}C, \ref{fig:profile}D). In Fig.~\ref{fig:2E0pres}F the spallation zone can be observed explicitly. The formation of rear cavity is accompanied by the tensile stress relaxation (the cavitation zone in Fig. 5F is represented by a light blue background, which corresponds to close-to-zero pressures). The neighboring regions at some distance from the opposite pole of the droplet are under large, but not sufficient to form a cavity, tensile stresses, and therefore the edges of the cavity have a corresponding bright blue color.

Relaxation of tensile stresses in the center of the droplet, as mentioned above, is accompanied by the formation of an ensemble of small cavitation bubbles. Their growth and further coagulation leads to the formation of a single cavity which eventually increases in size. At the moment $t^{*} = 12.925$ its radius is $x/R\sim0.5$ (Fig.~\ref{fig:frag}-1A) what corresponds to the size of the zone originally occupied by small bubbles. At the moment $t^{*} = 31.725$ the size of the cavity becomes comparable with the initial size of the droplet (Fig.~\ref{fig:frag}-1C) and continues to increase. At later stages, when the observed size of the droplet significantly exceeds its original size and its shell becomes sufficiently thin, the nature of the droplet deformation and fragmentation is determined by surface tension forces which are neglected in this work.

\subsection*{Effect of radiation intensity}

The additional mechanism of fragmentation at relatively high laser intensities is associated with spallation at the droplet's pole opposite to the irradiated one. Figure~\ref{fig:frag}A shows that a thin film is detached from the rear surface of the droplet as a result of spallation after shock-wave reflection. It should be noted that spallation takes place under a relatively small surface area near the droplet pole only, because the incident pressure pulse decreases with approaching the equatorial zone and is unable to initiate spallation. Thus, there is a qualitative analogy between the results obtained in the simulations and the experimental data: Irradiation of a liquid-metal tin droplet with a short high-intensity laser pulse leads to the formation of two spatially separated shells. According to our analysis, the first (central) shell emerges due to the growth of the cavity in the center of the droplet, whereas the second rear-side shell is formed as a result of spallation under the rear surface of the droplet.

The thickness of the central shell is minimal on the rear and front sides and much wider in the region of the equator (Fig.~\ref{fig:frag}-1C). Thus, further expansion of the rear-side shell should lead to its fragmentation. The thick equatorial ``ring'' should be more stable and fragment later. Such fragmentation sequence is confirmed by the experimental shadowgraphs presented in Figs.~\ref{fig:drop_decreas_int}, \ref{fig:Drop_low_int}. The formation of a ring is exhibited for shadowgraphs taken at the angle of $30^\circ$ to the horizontal projection of the laser beam (Fig.~\ref{fig:Drop_low_int}), what indirectly indicates that the emerging shell begins to break down at the poles of the droplet. The equatorial region is capable to maintain its continuity for a long time.

The laser intensity reduction leads to a gradual suppression of both cavitation and spallation at the center of the droplet and at the rear pole, respectively. The cavitation bubbles induced by a laser pulse with the twice lower intensity are formed within a smaller volume (Figs.~\ref{fig:frag}-2A--\ref{fig:frag}-2C). Moreover, the droplet shape evolution for decreasing intensities shown in Fig.~\ref{fig:frag} demonstrates a significant decrease in the cavity growth rate. At lower intensities the shells expansion rate becomes smaller what qualitatively agrees with the conclusions of the experiments. The intensity of cavitation and spall processes also decreases with the absorbed laser pulse energy. A decrease in the radiation intensity by a factor of~2 results in the damping of the shock wave amplitude to~$\sim0.8\un{GPa}$ near the rear pole of the droplet. This amplitude becomes insufficient to induce spallation during the release wave formation. Thus, there is a threshold value of the laser pulse intensity, below which the spall processes are completely suppressed. The further expansion of droplet and its fragmentation are determined only by cavitation processes near the center of the droplet. 

\subsection*{Similarity in droplet expansion}

The effect of the droplet size on expansion and fragmentation is studied using droplets with diameters $D = 2\un{\mu m}$ and $D = 4\un{\mu m}$. It should be noted that the problem of scaling effects is complex. First, the amplitude of the pressure pulse near its center can increase with increasing droplet diameter because the integral energy input rises. On the other hand, an increase in the size is accompanied by a large broadening of the initial pulse, thereby reducing the amplitude of the shock wave. Thus, there are two competing processes, one of which contributes to the increase in the amplitude of the shock wave, while the other, on the contrary, to its decrease.

Hydrodynamic similarity is realized for different droplet sizes if simultaneous scaling of spatial and temporal variables is performed in governing equations, boundary and initial conditions \cite{zeldovich2012physics}. Thus, solutions for droplets of different sizes will be similar if the depths of heating in them differ by the ratio of their diameters. In addition, it is necessary to similarly change the size of the SPH particles to keep the perfect similarity in simulation. Such scaling conserves the total deposited energy per droplet volume $E_{tot}/R^{3}$=const. Comparison of 1D velocity profiles obtained in the calculation of droplets with diameters $D=2\un{\mu m}$ and $D=4\un{\mu m}$ demonstrates their full similarity (see Appendix C). However, for the pulsed laser irradiation the heated depth is independent from the droplet size, which violates the perfect similarity just after the pulse, but the similarity is regained with time.

A pressure wave, generated as a result of instantaneous heating, begins to forget its original profile at a considerable distance from the thin surface layer of its formation. Due to the fact that the initial pressure pulse width is about 100 nm, this width will increase by an order of magnitude at a relatively small propagation distance because the large sound velocity dispersion in the given range of compression. Therefore, even if there is a difference in the widths of initial pressure pulses, it will be leveled with distance. Thus, the micrometer-sized pulse profiles will be asymptotically similar for the surface layers heated to the depths within a few hundred nano-meters, and their integral characteristics being preserved.
 
We will rely on the above-mentioned fact that solutions for different droplets are similar, provided the value of $E_{tot}/R^{3}$ is constant. In the first approximation, it can be assumed that an increase in the heated depth with increasing droplet size, and the input energy $E_{in,0}$ is fixed in Eq.~(\ref{eq:energI-3D}), produces a similar effect as an increase in $E_{in,0}$ while maintaining the heated depth. Thus, the scenario of the droplet fragmentation, whose size is $N$ times different, is expected to be identical if the intensity of the laser pulses that irradiate them is $N$ times different.

Figure~\ref{fig:vx-spall} presents a comparison of 1D velocity profiles obtained in the calculation of droplets with diameters $D=2\un{\mu m}$ (blue curve) and $D=4\un{\mu m}$ (red curve). In the calculation of a droplet with $D=4\un{\mu m}$, the maximum internal energy $E_{in,0}$ is doubled, and the depth of heating in both calculations is the same. The constancy of the heating depth in the calculations is justified, because its dependence on the laser pulse intensity is in fact insignificant. The spatial coordinate in Fig.~\ref{fig:vx-spall} is normalized to the radius of the droplet. Profiles are given for three different instants of times: before the formation of cavitation bubbles (Fig.~\ref{fig:vx-spall}A), as the shock front passes through the center of the droplet (Fig.~\ref{fig:vx-spall}B), and when the pressure pulse approaches the rear surface of the droplet (Fig.~\ref{fig:vx-spall}C).

Figure~\ref{fig:vx-spall} shows a good coincidence of the shock wave pressures and profiles of unloading tails. Because the spatial coordinate is normalized to the radius, then a good agreement of unloading waves indicates that the strain rate in the droplet with $D=4\un{\mu m}$ is two times greater than that in the droplet with $D=2\un{\mu m}$. The greatest difference is observed in the stretching region. It is seen in Fig.~\ref{fig:vx-spall}(A) that the negative stresses for a smaller droplet are larger in magnitude. Thus, it can be expected that in the case of smaller droplets, the formation of cavitation bubbles begins earlier. However, the dependence of the spall strength on the strain rate can slightly improve the coincidence. The shape of the profile remains similar when the pulse approaches the rear surface of the droplet; therefore, the conditions for the formation of a spall will be identical.

In general, a good coincidence of the profiles for droplets of different sizes suggests that the fragmentation described above for a droplet with $D=2\un{\mu m}$ will be similar for droplets of a larger size if $E_{tot}/R^{3}$ is kept constant. 

To quantitatively compare the calculation results with the experimental data, we estimate the energy of laser irradiation of the droplet at its total mass $3 E_{tot}/4\rho \pi R^3$. In the section describing the experimental setup, it is noted that the laser pulse intensity $I$ varied in the range $(0.4-8.0)\times10^{13}\un{W/cm^2}$. This corresponds to the specific energy $E_{exp}=3I\tau/4\rho R=(0.14-2.81)\times10^6\un{J/kg}$. At such laser intensities, the experimentally estimated velocity of the rear surface, which is carried away as a result of spallation, varies in the range from $100$ to $500\un{m/s}$ (Fig.~\ref{fig:vel}). In particular, at a specific energy $E_{exp}=1.408\times10^6\un{J/kg}$, the velocity of the rear surface is $300\un{m/s}$. Simulations of droplets with $D=4\,\mu\mathrm{m}$ give the same velocity of the rear surface at $E_{in,0}=7.5\times10^6\un{J/kg}$, which corresponds to the specific energy per total droplet mass $E_{sim}=0.176\times10^6\un{J/kg}$. The ratio $E_{sim}/E_{exp}\cong0.125$ can serve as an estimate of the absorption coefficient. The theoretically calculated absorption coefficient for liquid tin is $\sim20\,\%$\cite{siegel1976optical}, which agrees in order of magnitude with the given estimate. 

Figure~\ref{fig:vel} compares the rates of expansion of the front and rear shells, which were measured in experiments and calculations for different droplet irradiation energies, normalized per unit mass. In Fig. 9, circles denote the transverse expansion velocity of the front shell, and squares -- the rate of expansion of the rear shell in the longitudinal direction. It can be seen that the experimentally measured rate of expansion of the rear shell increases with increasing irradiation intensity much faster than in the front shell, which indicates different mechanisms of their formation. Estimated velocities show a similar behavior. There is good agreement between the results of the calculations and the experimental data under the condition of a constant absorption coefficient equal to 0.125 (Fig.~\ref{fig:vel}).

\section*{Conclusions} 

We have studied experimentally and numerically expansion and fragmentation of a liquid-metal droplet irradiated by a short laser pulse. The fast energy deposition heats and pressurizes material within a thin frontal layer of the droplet in the almost isochoric regime. Propagation of laser-generated pressure wave and physical mechanisms of fragmentation at various laser pulse energies are examined in detail. 

It is found in SPH simulation that fragmentation of a droplet is triggered by the increasing tensile stress produced in a rarefaction wave following the shock wave toward the center. As a result of rarefaction wave convergence, the central cavitation zone is formed in the droplet. For higher laser pulse energy the second cavity can be formed at the rear surface as a result of spallation of a thin rear-side layer after reflection of the shock wave from the free droplet boundary.

Two fragmentation regimes have been experimentally demonstrated. It has been shown that at a high intensity two spatially separated shells are formed, the expansion of which leads to fragmentation of the droplet. Only single spherical shell is formed in the low-intensity regime, at which the expansion and subsequent fragmentation of the droplet becomes more symmetrical in comparison with the high-intensity regime.

We have shown the existence of a critical intensity of the laser light below which the spall processes on the rear side of the droplet are completely suppressed. It has been shown in simulation that the fragmentation pattern is preserved for droplets of different sizes, provided that the absorbed energy of the laser pulse per total droplet mass remains constant.

\acknowledgments{
The works of S.Yu.G., V.V.Z., S.A.D, D.K.I, K.P.M. and N.A.I. were supported by the Russian Science Foundation grant 14-19-01599.
} 

\section*{Appendix A: Electron--ion exchange coefficients and electronic thermal conductivity}

Quantum molecular dynamics (QMD) simulations are performed for tin in the cubic super cell containing 64 atoms with periodic boundary conditions. The cube face size is $1.28\un{nm}$ which corresponds to the density of $6.05\un{g/cm^3}$. The simulation time step is 1~fs. There are 3 stages of simulation:
\begin{enumerate}
\item 1000 steps using the NVT thermostat with an increase in temperature from $300$ to $4000\un{K}$ in order to accelerate melting;
\item 1000 steps using the NVT thermostat with cooling from $4000$ to $1000\un{K}$;
\item 300 steps using the NVE ensemble (without a thermostat) to make sure that the final state is stable and remains liquid during the simulation ($2$-$3\un{ps}$).
\end{enumerate}
The effect of the semi-valence 4d electrons on the interatomic interaction is correctly taken into account by using the PAW pseudopotential which considers these electrons to be valence. The pseudopotential is a part of the library within the VASP software \cite{Kresse:PRB:1996,Kresse:CMS:1996}. The effects of electronic exchange and correlation are taken into account in the framework of the generalized gradient approximation (PBE) \cite{Perdew:PRL:1996}. The electronic structure is calculated for a single $\Gamma$-point using 544 blank electronic states per cell and the cutoff energy 260~eV.

The averaging over the last 60 configurations obtained in the stage 3 provides the electronic density of states (DoS) in liquid tin which is parabolic in the region near the Fermi energy. Similarly, DoS is computed for tin at equilibrium temperatures of 2000, 4000, and 8000~K without significant difference from the 1000~K result.

\begin{figure}[t]
\center{
\includegraphics[width=1.0\linewidth]{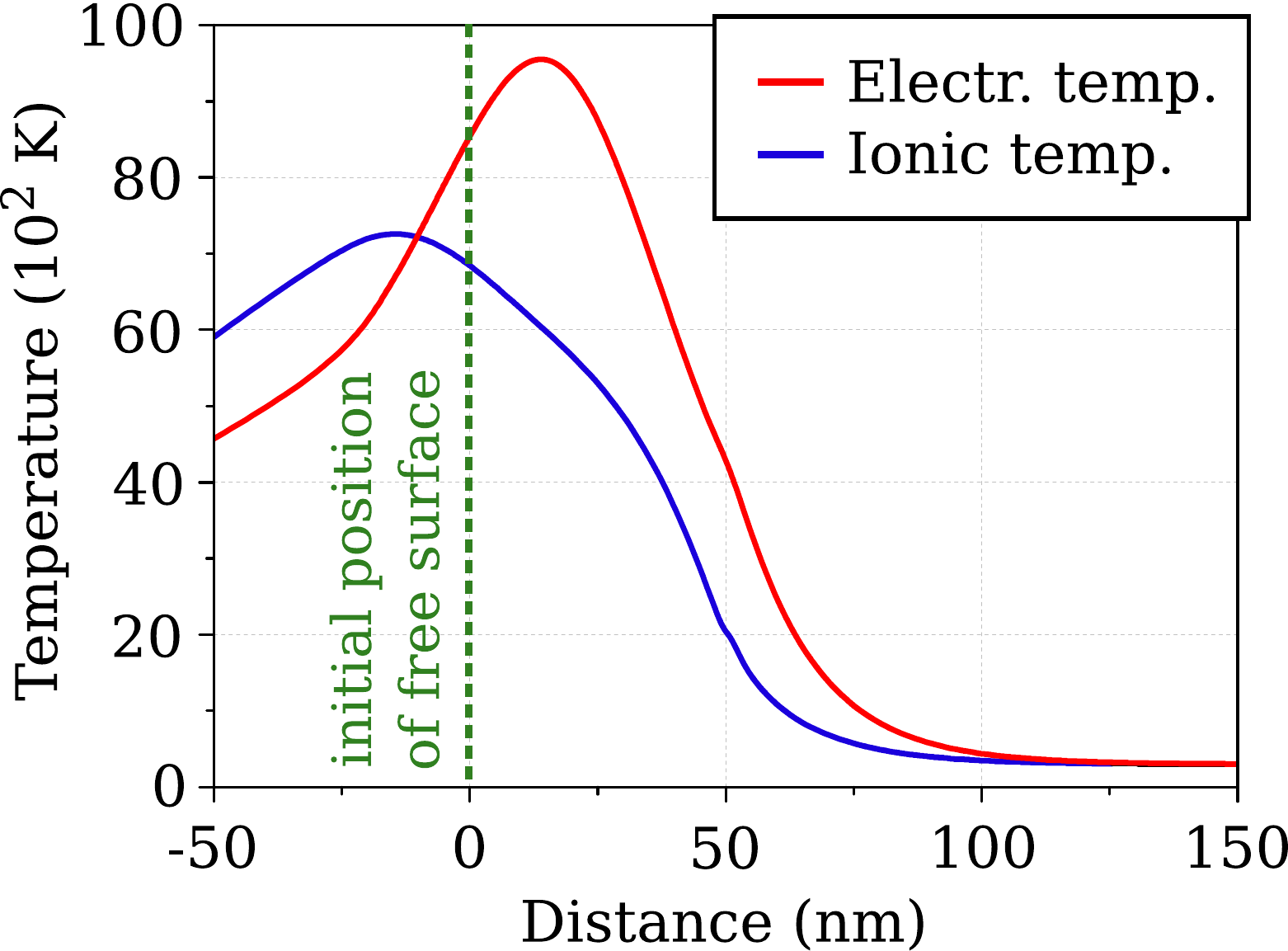} \\
\caption {\label{fig:Ti_Te} Temperature distribution of electrons and ions in 10 ps after irradiation.}
}
\end{figure}

In addition, we conducted a test to determine the effect of the electronic temperature on DoS. The unit cell of $\alpha$-tin with ion density of 6.05~g/cm$^3$ is considered. Electron temperatures to test are 10\,000, 20\,000, 45\,000, and 55\,000~K. The electronic exchange and correlation are taken again in the generalized gradient approximation (PBE) \cite{Perdew:PRL:1996}. Additional adjustments are the $21\times21\times21$ wave-vector mesh constructed by the Monkhorst--Pack algorithm, the plane-wave cutoff energy 300~eV, and the unfilled electronic states number per atom equal to 32. The performed tests do not demonstrate noticeable dependence of tin DoS on the electron temperature.

\begin{figure*}[t]
\center{
\includegraphics[width=0.9\linewidth]{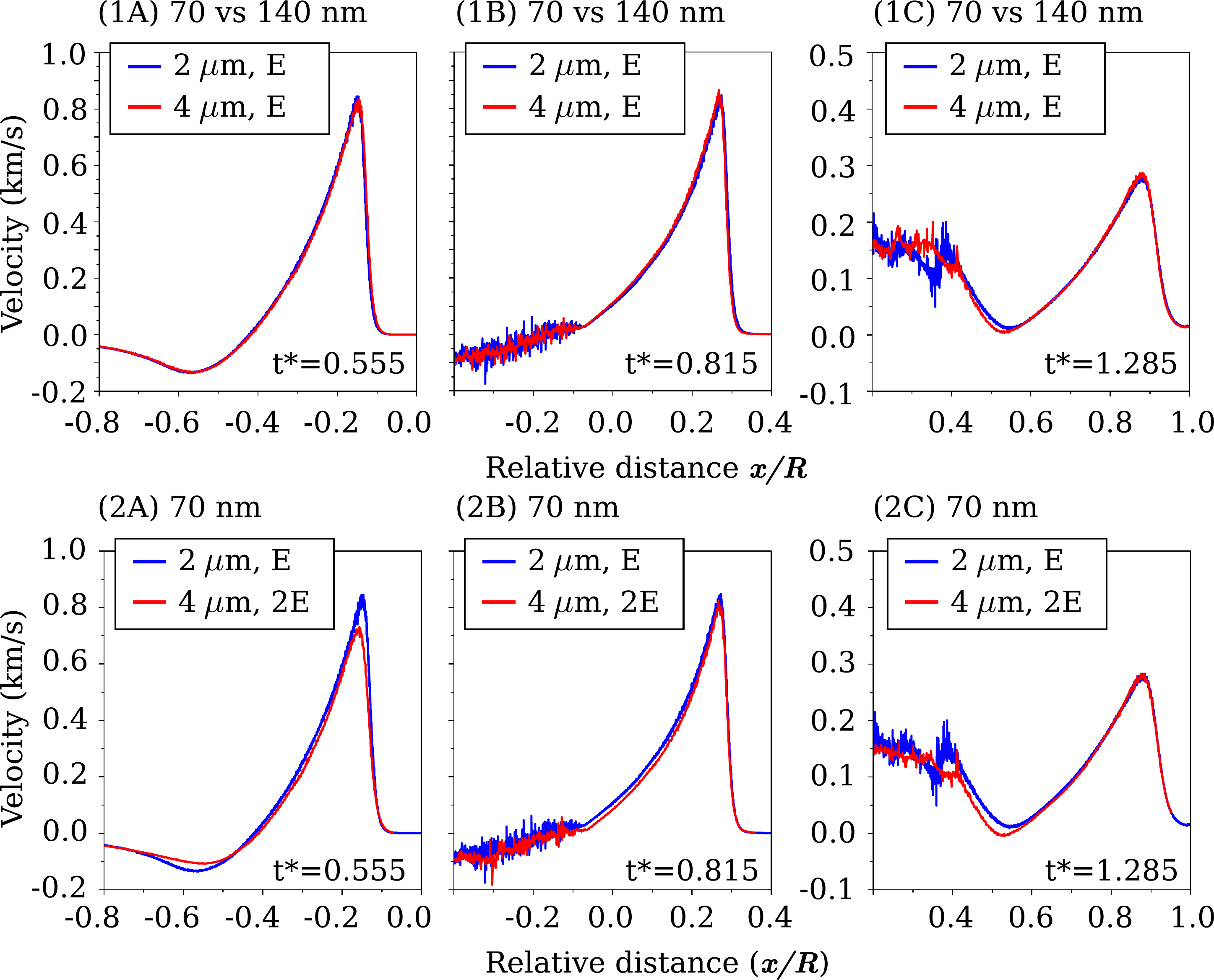} \\
\caption {\label{fig:analogy} (1A)--(1C): complete similarity of velocity profiles with a simultaneous change of all scales of length (droplet size, depth of heating, size of SPH particles) and time; (2A)--(2C): incomplete
(asymptotic) similarity of velocity profiles for droplets of various sizes while maintaining the depth of
heating and the specific absorbed energy per total mass of the drop.}
}
\end{figure*}

The two-parabolic approximation \cite{PetrovJETP:2013} which takes into account the presence of a semi-valence 4d band is used for further calculations. The sp-electron band is described by a parabola with the energy minimum of $-10.7$~eV and the effective electron mass of 0.83 in vacuum. The two-parabolic DoS of tin is used to obtain electronic thermodynamic characteristics. The expression for the electronic heat capacity (measured in 10$^5$\,J/m$^3$/K):
\begin{equation}
\label{1a}
C_e=1+4.64\times10^{-4}\,T_e+3.1\times10^{-9}\,T_e^{2},
\end{equation}
remains valid according to calculations with the two-parabolic DoS up to 100\,000~K. The role of d-electrons below $-20.5$~eV with respect to the Fermi energy manifested itself in a moderate increase in thermal conductivity at $T_e$ above 30\,000~K, which was already obtained for tungsten\cite{Levashov:2010}. The electron temperature $T_e$ in this and further expressions is given in K. The same applicability threshold corresponds to the obtained expressions the electron internal energy density (in GPa): 
\begin{equation}
\label{2a}
u_e=0.54-4.47\times10^{-6}\,T_e+3.0\times10^{-8}\,T_e^{2},
\end{equation}
and electron pressure:
\begin{equation}
\label{3a}
P_e=5.785\times10^{-5}\,T_e+2.462\times10^{-8}\,T_e^2.
\end{equation}

The effect of electron--electron collisions in the two-temperature electron thermal conductivity is taken into account in accordance with the modification \cite{Migdal:2015} of the approach \cite{PetrovJETP:2013}, where the loss of conducting properties is considered due to the increasing losses for thermoelectric phenomena at $T_e$ = 30\,000 -- 50\,000~K. The form, proposed by Lindhard, is used to describe the electronic screening. The final expression for the contribution of s--s collisions (in s$^{-1}$) to the total effective frequency of electron collisions has the form: 
\begin{equation}
\label{4a}
\nu_{ss}=\left( \sqrt{\left( 8\times10^{-5}\,T_e \right)^{2}+0.49} - 0.7\right)\times10^{15}.
\end{equation}
Using the Drude model,
The processing of the experimental data \cite{Savchenko,Patel} using the Drude model yielded the following estimation the effective frequency of electron--ion collisions of tin: 
\begin{equation}
\label{5a}
\nu_{si}=\frac{3\times10^{14}\,T_i}{36+0.09\,T_i}.
\end{equation}
Here, the temperature of tin ions $T_i$ is also given in K. The final expression for the electronic thermal conductivity (in W m/K) is also obtained using the Drude model. The s and p electrons are considered as charge and energy carriers used in the two-parabolic approximation:
\begin{equation}
\label{6a}
\kappa_e=\frac{C_s V_s^2}{3(\nu_{si}+\nu_{ss})}.
\end{equation}
The sp-electron velocity $V_s = 3\times$10$^{14}\,T_e$ (in SI units). 

The electron--phonon heat exchange is described using the approach \cite{PetrovJETP:2013}, where the experimental value of 2.4~km/s \cite{Hosokawa} is used for the sound speed in liquid tin. The approximate expression is:
\begin{equation}
\label{7a}
\alpha=0.41+1.91\times10^{-8}\times\,T_e+3.64\times10^{-11}\,T_e^2,
\end{equation}
which is given in units $10^{17}\un{W/K/m^3}$. 
Calculations are verified using the well-known Allen method \cite{Lin+LZ+Celli-2008} where $\lambda=0.6$ \cite{McMillan:1968} and the mean square of the oscillation frequency of acoustic phonons $\langle\omega^2\rangle$ are used for the electron-phonon coupling constant on the experimental data basis \cite{Chang:1986}. The resulting value of $\lambda \langle\omega^2\rangle$ used in the Allen formula is 19~MeV. The obtained value of electron-phonon heat transfer at electron temperatures up to 10\,000~K is 0.25$\times$10$^{17}\un{W/K/m^3}$. 

\section*{Appendix B: 2T hydrodynamics}

The system of equations for two-temperature hydrodynamics is solved numerically to estimate the skin layer $\delta R$ in tin. The nonequilibrium process of heating the electronic subsystem of the material and its further relaxation are taken into account. In the Lagrangian coordinates $dm=\rho dx$, the energy balance is divided into two independent equations for the electronic and ionic subsystems: 
\begin{equation}
\label{2T-Te}
\frac{\partial \epsilon _e}{\partial t}+ P_e \frac{\partial u}{\partial m}=\frac{\partial}{\partial m} \left( \kappa_e \rho \frac{\partial T_e}{\partial m} \right) - \frac{\alpha}{\rho} (T_e-T_i)+J_L, \\
\end{equation}
\begin{equation}
\label{2T-Ti}
\frac{ \partial\epsilon _i}{\partial t}+ P_i \frac{\partial u}{\partial m}=\frac{\alpha}{\rho} (T_e - T_i),
\end{equation}
where $\epsilon$, $P$, $T$ are the internal energy, the pressure and the temperature, respectively, for the electronic (subscript e) or ionic (subscript i) subsystems; $\alpha$ is the coefficient of electron--ion heat transfer, $\kappa_e$ is the coefficient of electronic thermal conductivity which is determined using QMD simulations. Material motion in one-dimensional hydrodynamics is described via the energy balance equations (\ref{2T-Te})--(\ref{2T-Ti}) which are supplemented by the equations of continuity momentum balance and the equation of state for the ionic post-system. The laser energy $J_L$ absorbed by a unit mass per unit time is represented in the form:
\begin{equation}
J_L=\frac{F_{abs}}{\tau _L \delta \sqrt{\pi} \rho} exp \left( - \frac{t^2}{\tau _L ^2} \right)exp \left( -\frac{x-x_0}{\delta} \right),
\end{equation} 
where $F_{abs}$ is the laser energy, absorbed by the unit of the irradiated surface, $\delta$ is the thickness of the skin layer, $x=x(m,t)$ and $x_0=x(m_0,t)$ are the trajectories of lagrangian particles with coordinates $m$ and $m_0$, $m_0$ is the lagrangian coordinate of the metal surface on which the light is incident, and $\tau_L$ is the laser pulse duration. 

A series of calculations is performed to estimate the heating depth. As a result, it varies from 60 to 80~nm over a wide range of fluences. Fig.~\ref{fig:Ti_Te} shows the distribution of the electron and ion temperatures 10~ps after irradiation by a 800-fs laser pulse with an energy density of 1~J/cm$^2$. 

\section*{Appendix C: Analysis of similarity}

A complete similarity in the evolution of irradiated droplets of various sizes is shown for two droplets with $D=2\,\mu \mathrm{m}$ and $D=4\,\mu \mathrm{m}$. The heated skin layers $\delta R$ are 70 and 140~nm for the first and the second case respectively. In addition to physical dimensions it is necessary to change the mesh size properly. Figures~\ref{fig:analogy}-1A--\ref{fig:analogy}-1C demonstrate the velocity profiles at different times for the droplets of different sizes (2 and 4~$\mu$m). The $x$ axis in the figures corresponds to the distance reduced by the droplet radius, and the origin corresponds to the droplet center. The obtained profiles almost completely coincide as expected.

Direct similarity in the evolution of droplets with a fixed skin layer depth with the same the specific absorbed energy $E/R^3$ is not observed. However, simulations demonstrate the asymptotic similarity what is shown in Figs.~\ref{fig:analogy}-2A--\ref{fig:analogy}-2C. The velocity profiles in the reduced coordinates for waves, approaching the droplet center, are very close. There is only a slight difference in the cavitation zone. Similarity takes place even when the pulse approaches the rear surface of the droplet (Fig.~\ref{fig:analogy}-2C).

\end{document}